\documentclass[onecolumn,showpacs,preprintnumbers,amsmath,amssymb,floatfix,elsart,prd]{revtex4}
\usepackage{graphicx}
\usepackage{dcolumn,epstopdf}
\usepackage{bm}
\usepackage{hyperref}
\hypersetup{colorlinks,citecolor=blue}

\newcommand{\beq}{\begin{equation}}
\newcommand{\eeq}{\end{equation}}
\newcommand{\bey}{\begin{eqnarray}}
\newcommand{\eey}{\end{eqnarray}}
\usepackage{color}

\begin{document}

\title{A note on shock wave in the dark matter medium}

\author{Mofazzal Azam}
\email{mofazzal.azam@gmail.com}
\affiliation{Visiting Professor, Centre for Cosmology and Science Popularisation (CCSP), SGT University, Gurugram 122006, India}

\author{M Sami}
\email{samijamia@gmail.com}
\affiliation{Centre for Cosmology and Science Popularisation (CCSP), SGT University, Gurugram 122006, India}

\author{Farook Rahaman}
\email{rahaman@iucaa.ernet.in}
\affiliation{Department of
Mathematics, Jadavpur University, Kolkata 700032, West Bengal, India.}

\date{\today}

\begin{abstract}
Employing Newton's stellar balance equation and using the flat rotation velocity of satellite galaxies, we have found the velocity of sound in the dark matter medium. What is interesting is that the velocity of satellite galaxies is very much larger than the velocity of sound in the dark matter medium. This indicates that there will be shock wave in the dark matter medium due to the supersonic movement of satellite galaxies.

\end{abstract}

\maketitle

Dark matter is an active field of research in astrophysics and cosmology. From astronomical observations, there does not exist any doubt about the existence of dark matter effects. However, the true physical nature of the dark matter is still unknown. In cosmology, the $\Lambda$CDM model describes the cosmological observation and structure formation in the universe fairly well. But this is based on large scale global observation. Local observation of the expansion of the universe does seem to fully reconcile with it. Infrared modification of gravity remains a possibility.

 To know the spacetime geometry of galactic halo comprising
the dark matter is a very pertinent issue.  in the year 1976, Lynden-Bell \cite{1} revealed a number of satellite galaxies
around the Milky Way.  It is argued in recent time that satellite galaxies around the Milky Way gather close to the
galactic poles in the vast polar structure.  It is observed that the motion of satellite galaxies around Andromeda
Galaxy are non-isotropic \cite{2}.
It is now well establish that the velocities of satellite galaxies around the central massive galaxy do not fall off with distance- it is in fact constant independent of distance \cite{3,4}. This has come to be known as flat rotation curve. This observation can be understood in two different ways:

 (1)  There is  infrared modification (large distance/low density/low acceleration limit) of gravity. Solution of Poisson equation for a mass distribution, gives the gravitational potential as the usual inverse distant dependent function plus a constant. This constant is chosen to be zero by choosing a zero boundary condition at infinity. However, if we choose instead of the constant, a distant dependent logarithm, then this term would behave almost as a constant growing very  slowly with the increase of distance.

 A choice of
\[\phi (R) =  -\frac{GM}{R_0}  \ln\left(\frac{R}{R_0}\right), \] does the job, and we have the
flat rotation curve. Here $R_0$ is the dimension of the
central high density region of the galaxy, and
$R_0<<R$. This is in fact a version of Milgrom-Beckenstein theory of gravitation popularly known
as MOND \cite{5}.

(2) The infrared modification can also be achieved by assuming that central galaxy is surrounded by dark matter with inverse square density profile. This may look very simple and a very natural choice  but one should remember that the inverse square density profile have to meet some thermodynamic conditions for stability. We show in this letter that such a requirement leads to the equation of state of dark matter with velocity of sound much smaller than the velocity of satellite galaxies in the flat rotation curve. The supersonic velocities of the satellite galaxies fulfils the condition for shock wave generation in the dark matter medium.

The velocity, $ v $   corresponding to the flat rotation curve of large galaxies such as the Milky Way and Andromeda is
 of the order of
  \[  300 km/sec  = 3\times10^7 cm/sec = 10^ {-3} ( in~ c = 1~ unit ) .\]

Let us assume dark matter density profile as \cite{6},
\begin{equation}\rho = \frac{\sigma }{ r^2}, \end{equation} where  $\sigma $ is a constant, whose unit is gm/cm in CGS system. \\

Therefore, the mass of sphere of radius   r is

\begin{equation} M(r) = \int \rho \times d^3x =\int \frac{\sigma }{ r^2} 4\pi  r^2 dr =4\pi \sigma r.\end{equation}

Velocity of a satellite galaxy at distance $r$ is given by,
\[ \frac{v^2}{r} =\frac{GM(r)}{r^2}  =\frac{G\times 4 \pi \sigma r}{r^2} =\frac{4 \pi \sigma G}{r}.\]
Thus,
\begin{equation} v^2 = 4 \pi \sigma G.\end{equation}
 Note that the equation above can also be written as,
\[ \frac{v^2}{r} = - 4 \pi \sigma G  \nabla \phi, \]
where, \[ \phi = - \ln \frac{r}{r_0}.\]
Thus the dark matter with inverse square density profile does the job of infrared/long range modification of gravity.

From the equation, $v^2=4 \pi \sigma  G$, we can find the value of $\sigma $,  for large galaxies.

\[\sigma =\frac{v^2}{4 \pi G }
                    = \frac{(3\times10^7)^2}{4 \pi \times 6.68\times10^{-8}}
                 \]
                    i.e.,
               \begin{equation} \sigma
                    =1.09 \times 10^{21} gm/cm .   \end{equation}

Taking this   value of $ \sigma$  and $r  = 15  ~Kpc  = 4.5\times 10^{22}  cms $
for the radius of central bulge of large galaxy like Milky Way, we find the mass contained within this radius to be of the order of  $3.1\times10^{11}   M_\odot$  which is a reasonable value.

Flat rotation curve of satellite galaxies around the heavier galaxies is in the infrared region of the gravity. Thus for the stability of dark matter sphere around the central galaxy, we can use Newton's equation for stellar balance.
The equation of stellar balance is,

  \begin{equation}
-r^2 \frac{dP}{dr} = G \times M(r) \times \rho(r)       \end{equation}
 Using the values of $M(r)$     and $\rho(r)$    we get,
\begin{equation}  \frac{dP}{dr} = -\frac{4 \pi G \sigma^2}{r^3}.\end{equation}
Solving this equation, we get

\[ P = \frac{2 \pi G \sigma^2}{r^2}
 =2\pi G \sigma \times\ \rho\]
Thus
\begin{equation} c^2_s = \frac{dP}{d\rho} = 2\pi G \sigma,\end{equation} where $c_s$ - is the velocity  of sound in the medium.\\

With \[ \sigma =1.09\times 10^{21}  gm/cm ,~~~ c_s=(2\pi \sigma G)^{1/2} \approx 214 ~ km/sec .\]  But the velocity of satellite galaxy around the central galaxy is, \[ v=(4\pi \sigma G)^{1/2}\approx 300 ~ km/sec. \]
Thus the velocity of satellite galaxy is larger than the velocity sound in the dark matter medium.
This supersonic velocity fulfils the condition for shock wave generation in the dark matter medium.

Note that in the analysis above, we could have used Planckonion type density parametrization \cite{7},

\[    \rho = \frac{u^2}{8 \pi Gr^2 }   ~ , ~  u<1,~~in~ the ~unit~ c = 1.\]
Thus, \[ \sigma = \frac{u^2}{8 \pi G } ~~ ;~ c_s=u/2~; ~~ v=\frac{u}{\sqrt{2}}. \]

We will make use of the  Planckonion type parametrization of matter density profile for the stability analysis of dark matter sphere in general relativity. In this case, it is easier to use the TOV equation,

\begin{equation}-r^2 \frac{dP}{dr} \left(1-\frac{2 G M(r)}{r}\right)
= G(P+\rho)[M(r)+4 r^3 \rho].\end{equation}
Here we have chosen $ c=1.$ \\
Now we use the old parametrization of density , $\rho=\frac{\sigma}{r^2 }$ , for some simplifications.
\[
\left(1-\frac{2 G M(r)}{r}\right)=\left(1-8 \pi G \sigma  \right) =  1-u^2. \]
\[ M(r)+4\pi r^3 P= (4 \pi \sigma r +4 \pi r^3 P)
= 4 \pi r^3 \left[\frac{\sigma}{r^2}+P\right]  =
4 \pi  r^3 (\rho +P).\]

The TOV equation now takes the form

  \begin{equation}  -r^2 \frac{dP}{dr}(1-u^2)=4 \pi G r^3 (\rho +P)^2.\end{equation}

Taking, \[ P=\omega \rho ,   ~~ where ~\omega =c_s^2,~~   \rho =\frac{u^2}{8\pi Gr^2}, \]
we get ,
 \begin{equation} \omega  (1-u^2)=\frac{u^2}{4} (1+w)^2 . \end{equation}

When $u\rightarrow 1$ (in c=1 unit), we have planckonian , $w=-1$  and the Equation  state is   \[P =-\rho.\]

For the case of dark matter \[ u<<1,~~  w<<1 .\]
Thus dropping terms $wu^2 , u^2w^2 ,u^2w$,  we get from TOV equation
\[ \omega =\frac{u^2}{4}~ ; ~\omega =c^2_s.\]
Thus we get our non-relativistic relation
\[ c^2_s = \frac{u^2}{4}\]
$\Rightarrow$\[  c_s = \frac{u}{2}.\]
 $\Rightarrow$\[  v =  \sqrt{2}c_s = \frac{u}{\sqrt{2}}.\]

As expected, in the small velocity limit of the TOV equation we obtain the same result as before.
We have demonstrated in this letter that the velocity of satellite galaxies around the central galaxy  is very much larger than the velocity of sound in the intervening dark matter medium. This yields in interesting result that there will be shock wave in the dark matter medium.

 In future, apart from the gravitational interactions lesson in the
galactic halo region,  one can quest for a cross- verification of the  presence of
dark matter   by conducting   several  local gravitational events  such
as gravitational lensing, gravitational time delay,  shadow cast  etc.

\section*{Acknowledgments}
  FR would like to thank the authorities of the Inter-University Centre for Astronomy and Astrophysics, Pune, India for providing research facilities.


\begin{thebibliography}{99}



\bibitem{1} Lynden-Bell D , MNRAS, 1976, vol. 174 pg. 695.
\bibitem{2} J Lopez et al, Phys.Rev.D 103 (2021) 8, 083535
\bibitem{3} J.Einasto, A.Kaasik and E.Saar,  Nature,250 309(1974)
\bibitem{4} V.C.Rubin,  Science,220 1339 (1983)
\bibitem{5} Bekenstein, J., Milgrom, M., 1984. Ap. J., 286, 7
\bibitem{6} F Rahaman, K K  Nandi,  A Bhadra, M Kalam, K Chakraborty, Phys.Lett.B 694 (2011) 10-15, e-Print: 1009.3572[gr-qc]
\bibitem{7} M Azam, F Rahaman, M Sami, J Bhatt, Mod.Phys.Lett.A 34 (2019) 33, 1950268







\end{thebibliography}
\end{document}